# COVID-19 Infection Localization and Severity Grading from Chest X-ray Images

Anas M. Tahir [1*], Muhammad E. H. Chowdhury[1*], Amith Khandakar[1], Tawsifur Rahman[1], Yazan Qiblawey[1], Uzair Khurshid[1], Serkan Kiranyaz[1], Nabil Ibtehaz[2], M Shohel Rahman[2], Somaya Al-Madeed[3], Khaled Hameed[4], Tahir Hamid[5], Sakib Mahmud[1], Maymouna Ezeddin[1]

[1]Department of Electrical Engineering, Qatar University, Doha-2713, Qatar; a.tahir@qu.edu.qa (AMT), mchowdhury@qu.edu.qa (MEHC), amitk@qu.edu.qa (AK), tawsifurahman.1426@gmail.com (TR), yazan.qiblawey@qu.edu.qa (YQ), uk1506741@qu.edu.qa (UK), mkiranyaz@qu.edu.qa (SK), sm1512633@qu.edu.qa (SM), maymouna@qu.edu.qa (ME)

[2]Department of Computer Science and Engineering, Bangladesh University of Engineering and Technology, Dhaka-1205, Bangladesh; 1017052037@grad.cse.buet.ac.bd (NI), msrahman@cse.ac.buet.bd (MSR)

[3]Computer Science and Engineering, Qatar University, Doha-2713, Qatar; s_alali@qu.edu.qa (SAM)

[4]MD, Reem Medical Center, Doha, Qatar, dr.khalid@reemmedicalcenter.com (KH)

[5]Consultant cardiologist in Hamad General Hospital and Weill Cornell Medicine - Qatar, Doha, Qatar, tahirhamid76@yahoo.co.uk (TH)

*Corresponding authors: Muhammad E. H. Chowdhury (mchowdhury@qu.edu.qa)

Anas M. Tahir (a.tahir@qu.edu.qa)

**Abstract**

Coronavirus disease 2019 (COVID-19) has been the main agenda of the whole world, since it came into sight in December 2019 as it has significantly affected the world economy and healthcare system. Given the effects of COVID-19 on pulmonary tissues, chest radiographic imaging has become a necessity for screening and monitoring the disease. Numerous studies have proposed Deep Learning approaches for the automatic diagnosis of COVID-19. Although these methods achieved astonishing performance in detection, they have used limited chest X-ray (CXR) repositories for evaluation, usually with a few hundred COVID-19 CXR images only. Thus, such data scarcity prevents reliable evaluation with the potential of overfitting. In addition, most studies showed no or limited capability in infection localization and severity grading of COVID-19 pneumonia. In this study, we address this urgent need by proposing a systematic and unified approach for lung segmentation and COVID-19 localization with infection quantification from CXR images. To accomplish this, we have constructed the largest benchmark dataset with 33,920 CXR images, including 11,956 COVID-19 samples, where the annotation of ground-truth lung segmentation masks is performed on CXRs by a novel human-machine collaborative approach. An extensive set of experiments was performed using the *state-of-the-art* segmentation networks, U-Net, U-Net++, and Feature Pyramid Networks (FPN). The developed network, after an extensive iterative process, reached a superior performance for lung region segmentation with Intersection over Union (IoU) of 96.11% and Dice Similarity Coefficient (DSC) of 97.99%. Furthermore, COVID-19 infections of various shapes and types were reliably localized with 83.05% IoU and 88.21% DSC. Finally, the proposed approach has achieved an outstanding COVID-19 detection performance with both sensitivity and specificity values above 99%.

## 1 Introduction

The novel coronavirus 2019 (COVID-19) is an acute respiratory syndrome that has already caused over 2.6 million causalities and infected more than 117 million people, as of March 11, 2021 [1]. The business, economic, and social dynamics of the whole world were affected. Governments have imposed flight restrictions, social distancing, and increasing awareness of hygiene. However, COVID-19 is still spreading at a very rapid rate. The common symptoms of coronavirus include fever, cough, shortness of breath, and pneumonia. Severe cases of coronavirus disease result in acute respiratory distress syndrome (ARDS) or complete respiratory failure, which requires support from mechanical ventilation and an intensive-care unit (ICU). People with a compromised immune system or elderly people are more likely to develop serious illnesses, including heart and kidney failures and septic shock [2].

Intuitively, reliable detection of COVID-19 disease has the utmost importance. However, the diagnosis procedures are not straightforward, as the common symptoms of COVID-19 are generally indistinguishable from other viral infections [3, 4]. Currently, the primary diagnostic tool to detect COVID-19 is reverse-transcription polymerase chain reaction (RT-PCR) arrays, where the presence of Severe Acute Respiratory Syndrome Related Coronavirus 2 (SARS-CoV-2) Ribonucleic acid (RNA) is tested on collected respiratory specimens from the suspected case [5, 6]. However, RT-PCR arrays have a high false alarm rate caused by sample contamination, damage to the sample, or virus mutations in the COVID-19 genome [7, 8]. Therefore, several studies suggested using chest computed tomography (CT) imaging as a primary diagnostic tool since it shows higher sensitivity values compared to RT-PCR [9, 10]. Besides, several studies [9-11] suggest performing CT as a secondary test if the suspected patients with shortness of breath or other respiratory symptoms showed negative RT-PCR findings. Despite the superior performance, CT scans pose difficulties and certain limitations. Their sensitivity is limited for early COVID-19 cases, slow in image acquisition, less applicable, and bear high costs. On the other hand, X-ray imaging is a cheaper, faster, and readily available method, where the body gets exposed to smaller amounts of harmful radiation compared to CT [12]. Chest X-ray (CXR) imaging is widely used as an assistive diagnostic tool in COVID-19 screening, and it is reported to have high potential prognostic capabilities [13].

The majority of early COVID-19 cases show similar features on radiographic images, including bilateral, multi-focal, ground-glass opacities with posterior or peripheral distribution, mainly in the lower lung lobes, while it develops to pulmonary consolidation in the late stage [14, 15]. Even though chest radiographs can help in the early screening of the suspected case, the images of several viral pneumonia are similar. They show a high overlap with other inflammatory lung diseases. Therefore, it is difficult for medical doctors to distinguish COVID-19 infection from other viral pneumonia. Hence, this symptom similarity can lead to the wrong diagnosis in the current situation, which may cause delayed treatment or even cost human lives.

The tremendous development in Deep Learning techniques in recent years led to state-of-the-art performance in several Computer Vision tasks, such as image classification, object detection, and image segmentation. This breakthrough led to increased utilization of AI-based solutions in various life fields, including biomedical health problems and complications. Specifically, Convolutional Neural Network (CNN) has been proven extremely beneficial in several biomedical imaging applications, such as skin lesion classification [16], brain tumor detection [17], breast cancer detection [18], and lung pathology screening [19, 20]. Deep Learning techniques on chest X-ray images are gaining popularity with the availability of deep CNNs, showing promising results in various applications. Rajpurkar et al. [21] proposed the CheXNet network, one of the top-performing architectures for CXR, by training Densenet121 on the ChestX-ray14 dataset [22], the largest public CXR dataset with over 100 thousand X-ray images for 14 different pathologies.

Rahman et al. [23] investigated several pre-trained CNNs to classify the CXR images as having manifestations of pulmonary tuberculosis (TB) or as healthy. The proposed model was trained over a dataset of 3,500 infected and 3,500 Normal CXR images. A high detection performance was achieved by the best performing model, DenseNet201, with 98.57% sensitivity and 98.56% specificity.

Recently, many studies have reported Deep Learning approaches to automate COVID-19 detection from chest X-ray images [24-34]. They have reported high detection performance for the disease; however, they also present certain issues and drawbacks. First of all, all of them have used a limited amount of COVID-19 data, e.g., the largest dataset includes only a few hundred CXR samples. As mentioned earlier, such a data scarcity yields a lack of proper evaluation, and thus it is difficult to generalize their results in practice. Moreover, they only aimed for COVID-19 detection and/or classification among other types without further assessment and localization. Due to these issues, their usability and robustness for a clinical usage will be very limited.

On the other hand, few studies [33, 34] considered lung segmentation as the first stage in their detection system. This ensures reliable decision-making in the classification phase and guards the network against irrelevant features from non-lung areas, such as heart, bones, background, or text. However, the previous segmentation approaches were trained on a mixture of medium and high-quality CXR images, mainly from Montgomery [35] and Shenzhen [36] CXR lung mask datasets which combinedly creates 704 X-ray images for Normal and TB cases. Therefore, the segmentation performance degrades in unseen scenarios such as severe COVID-19 cases or low-quality images with poor signal-to-noise (SNR) levels. Hence, lung areas can be partially or incompletely segmented for severe COVID-19 infections such as bilateral consolidation or fluid accumulation at lower-lung lobes, which degrades the classification performance. Therefore, creating a large benchmark CXR dataset with ground-truth lung segmentation masks is of high importance, and will help the research community to provide a more reliable detection system for COVID-19 and other lung pathologies.

Along with COVID-19 detection, infection localization is another crucial task that helps in evaluating the status of the patient and in the treatment process [37]. Therefore, several studies utilized class activation maps which are generated from Deep Learning models trained for COVID-19 classification tasks to localize infected lung regions. Those localized regions are potential signatures for COVID-19. However, more precise and reliable localization can be provided by ground-truth infection mask from expert radiologists. Therefore, Degerli et al. [38] proposed a novel approach for COVID-19 infection map generation by compiling a COVID-19 dataset consisting of 2,951 CXR images with annotated ground-truth infection segmentation masks. Several encode-decoder (E-D) CNNs were trained and evaluated on the generated dataset, where the best performing network achieved an 85.81% f1-score for infection localization. However, their proposed approach is limited only to infection localization. Therefore, there is room to revisit the problem with both lung and infection segmentation models to both localize and quantify infection regions by computing the overall percentage of infected lungs. This can help medical doctors to quantify the severity and track the progression of COVID-19 pneumonia.

In this work, in order to overcome the aforementioned limitations and challenges, we have accomplished the following objectives:

- The largest COVID-19 benchmark dataset, namely COVID-QU, has been created with 11,956 COVID-19, 11,263 Non-COVID, and 10,701 Normal (healthy) chest X-ray (CXR) images. This will not only provide the most reliable evaluation ever performed for COVID-19 detection, localization, and quantification; it will also help to investigate the *state-of-the-art* deep network models more reliably and accurately.

- For the first time, the ground-truth lung segmentation masks for the entire COVID-QU dataset have been created by using a novel human-machine collaborative approach that significantly reduces human labor to annotate the images. Both the dataset and the ground-truth masks will be released along with this study as public benchmark dataset. We believe that COVID-QU will be extremely beneficial for researchers, doctors, and engineers around the world to come up with innovative solutions for the early detection of COVID-19 with the help of the large benchmark COVID-19 CXR images with their ground-truth lung masks.

- Furthermore, we have investigated the three *state-of-the-art* image segmentation architectures, U-Net, U-Net++, and Feature Pyramid Networks (FPN) with different backbone encoder structures starting from shallow to deep structures: ResNet18, ResNet50, DenseNet121, DenseNet161, and InceptionV4 for both lung and infection segmentation tasks and thus found out which model is the best for each task accomplished.

- Finally, we have proposed a novel and robust system for lung segmentation and COVID-19 localization with infection quantification from CXR images. This is a crucial accomplishment for a reliable diagnosis and assessment of the disease with the highest accuracy ever reached.

## 2 The Benchmark COVID-QU Dataset

In this section, we will first show the data compilation process; then, we will present the proposed approach for ground-truth lung mask generation.

*2.1  Data Compilation*

Due to the emerging nature of the pandemic, initially, little efforts have been made by highly infected countries on sharing clinical and radiography data publicly. Therefore, a group of researchers from Qatar University (QU) and Tampere University (TU), have created two datasets, the so-called COVID-QU [39] and QaTa-Cov19 datasets [38]. The COVID-QU dataset consists of 3,616 COVID-19, 8,851 Non-COVID cases, and 6,012 Normal cases. While QaTa-Cov19 dataset includes 2,951 COVID-19 CXR along with their ground-truth infection masks. Gradually, more X-rays have become publicly available. Hence, we extended COVID-QU to include over 33,000 CXR images, from three different classes:

1) 11,956 COVID-19
2) 11,263 Non-COVID infections (viral or bacterial pneumonia)
3) 10,701 Normal (healthy)

In this study, only posterior-to-anterior (PA) or anterior-to-posterior (AP) chest X-rays were considered as this view of radiography is widely used by the radiologist. This dataset was created by utilizing numerous publicly available datasets and repositories, all of which are scattered, and with varying formats. Authors ensured the quality of the provided information; duplicates, extremely low-quality, and over-exposed images were identified and removed in the preprocessing stage. Consequently, the dataset encapsulates images of high interclass dissimilarity with varying resolution, quality, and SNR levels, as shown in Figure 1.

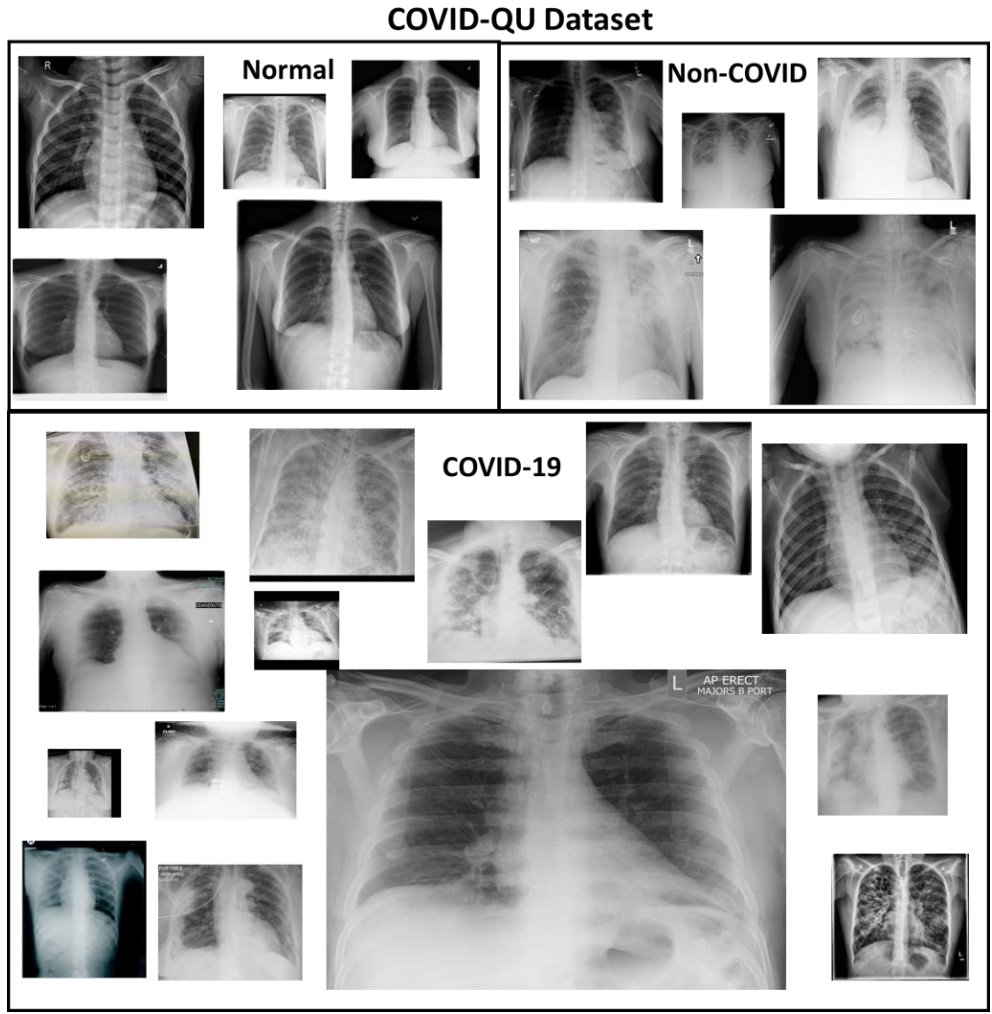

Figure 1. Sample chest X-ray images from the COVID-QU dataset for COVID-19, Non-COVID, and Normal classes.

Details of different data sources are given below:

**COVID-19 CXR dataset:** The dataset contains 11,956 positive COVID-19 CXR images: 10,814 images are collected from BIMCV-COVID19+ dataset [40], 183 images from a German medical school [41], 559 X-ray image from SIRM, Github, Kaggle, and Tweeter [42-45], and 400 X-ray images from another COVID-19 chest X-ray repository [46].

**RSNA CXR dataset (Non-COVID infections and Normal CXR):** RSNA pneumonia detection challenge dataset [47] consists of 26,684 chest X-ray images, where 8,851 images are normal, 11821 are abnormal, and 6012 are lung opacity images. All images are in DICOM format. In this study, we used 8851 normal and 6012 lung opacity X-ray, where lung opacity images are used as a Non-COVID class.

**Chest-Xray-Pneumonia dataset:** This is a Kaggle dataset [48] that encapsulates 1,300 viral pneumonia, 1,700 bacterial pneumonia, and 1,000 normal X-rays. In this study, viral and bacterial pneumonia are considered as Non-COVID-19 class.

**PadChest dataset:** PadChest [49] dataset comprises more than 160,000 X-ray images from 67,000 patients that were collected and reported by radiologists at Hospital San Juan (Spain) from 2009 to 2017. In this study, we used 4,000 normal, and 4,000 pneumonia/infiltrate (Non-COVID-19) cases.

**Montgomery and Shenzhen CXR lung masks dataset:** This dataset consists of 704 CXR images with their corresponding lung segmentation masks. It was used as initial ground truth masks to train the lung segmentation model in the first stage of the proposed human-machine collaborative approach. The dataset was acquired by Shenzhen Hospital in China [36], and the tuberculosis control program of the Department of Health and Human Services of Montgomery County, MD, USA [35]. Montgomery dataset consists of 80 normal and 58 tuberculosis CXR with lung segmentation masks. While Shenzhen dataset compromises 326 normal and 336 tuberculosis CXR, where 566 out of 662 CXR are provided with their corresponding masks.

**QaTa-Cov19 CXR infection mask dataset** [38]**:** This dataset was created by a research group from Qatar University and Tampere University. It consists of nearly 120K CXR images, including 2,913 COVID-19 images with their corresponding ground-truth infection masks. In this study, the ground-truth infection masks were used to train and evaluate the infection segmentation models.

*2.2   Collaborative Human-Machine Segmentation Approach for Lung Ground-Truth Mask Generation*

Recent advancements in Deep Learning techniques led to remarkable success; however, supervised Deep Learning approaches require large and annotated data for training. Otherwise, scarcity of data degrades their performance, resulting in poor generalization capabilities. Nevertheless, the process of producing ground truth segmentation masks is an exhaustive task, where human experts need to delineate pixel-wise masks. This is not only a demanding task for a human; however, the resultant segmentation masks will suffer from the varying subjectivity and hand-crafting levels of the human annotators.

To overcome this issue, a *collaborative* human-machine segmentation approach is proposed to accurately produce the ground-truth lung segmentation masks for CXR images.

The human-machine collaborative approach was performed in four main stages.

**Stage I (Initial Training):**
In the first stage, three variants of the U-Net [50] segmentation model, are trained on 704 CXR

images and ground-truth lung masks publicly available from Montgomery and Shenzhen dataset mentioned previously. The ground-truth CXR lung masks are referred to as the CXR lung mask repository in Figure 2, and it is enlarged throughout the mask creation process. Next, the best performing network in terms of Dice Similarity Coefficient (DSC) is selected as the main network for Stage II, which is referred to as the CXR-Segmentation network in Figure 2.

**Stage II (Collaborative Evaluation):**

In the second stage, an iterative training is utilized to create lung masks for a subset of 3,000 CXR samples (~10% of the full dataset) that well represent the diversity of the COVID-QU dataset. Firstly, A subset of 500 samples is selected and inferred using the CXR-Segmentation model. The predicated lung masks are then evaluated by researchers as, "accept", "reject", "unsure", or "exclude". Accepted masks that accurately cover the lung areas are added into the CXR-lung-mask-repository. Rejected masks either miss certain parts of the lung or include irrelevant parts. Those rejected masks are then manually modified by the researchers, and the corrected masks are finally added to the CXR-lung-mask-repository. The "unsure" masks are those severe cases with highly infected areas. They are usually consolidations or fluid accumulation at lower lung lobes with a whitish color, which makes them indistinguishable from neighboring organs. The unsure masks are first assessed by MDs; then, researchers adjust the masks based on their recommendations. Finally, the "excluded" masks are the ones where the quality is extremely bad for a proper lung segmentation. Finally, the CXR-Segmentation (best-performing) network is re-trained on the extended mask dataset. Then the second subset of 500 samples is selected, and the steps of Stage II are repeated. This process is repeated until generating ground-truth masks for 3,000 CXR samples is completed.

**Stage III (Collaborative Selection):**

In the third stage, six deep segmentation networks from the models of U-Net [50], U-Net++ [51], and FPN [52], are trained using the 3,000 ground-truth masks generated in Stage II by the proposed approach. The trained networks are used to predict segmentation masks for the rest of the COVID-QU dataset, which is around 30,900 unannotated samples (~90% of the full dataset). Among the six predictions, researchers select the best one as the ground-truth or deny if none of the masks segments the lung properly. The latter is a minority case that included less than 5% of unannotated data. The most selected network is considered as the best-performing network and used for a new training with the extended masks repository. The denied cases are then inferred by the main segmentation network and evaluated manually following the steps in Stage II. As a result, the ground-truth masks for 33,920 CXR images are gathered to construct the benchmark COVID-QU lung masks dataset.

**Stage IV (Final Verification):**

In the final stage, a final verification is performed by MDs on randomly selected 6,788 CXR samples (20% of the full dataset) that well presents the diversity of the COVID-QU dataset. The samples are selected from COVID, Non-COVID, and Normal classes, with different resolution, quality, and SNR levels. In this study, the verified subset (20%) was considered as a test set for all the experimental evaluations, while the remaining data (80%) were considered as train and validation sets.

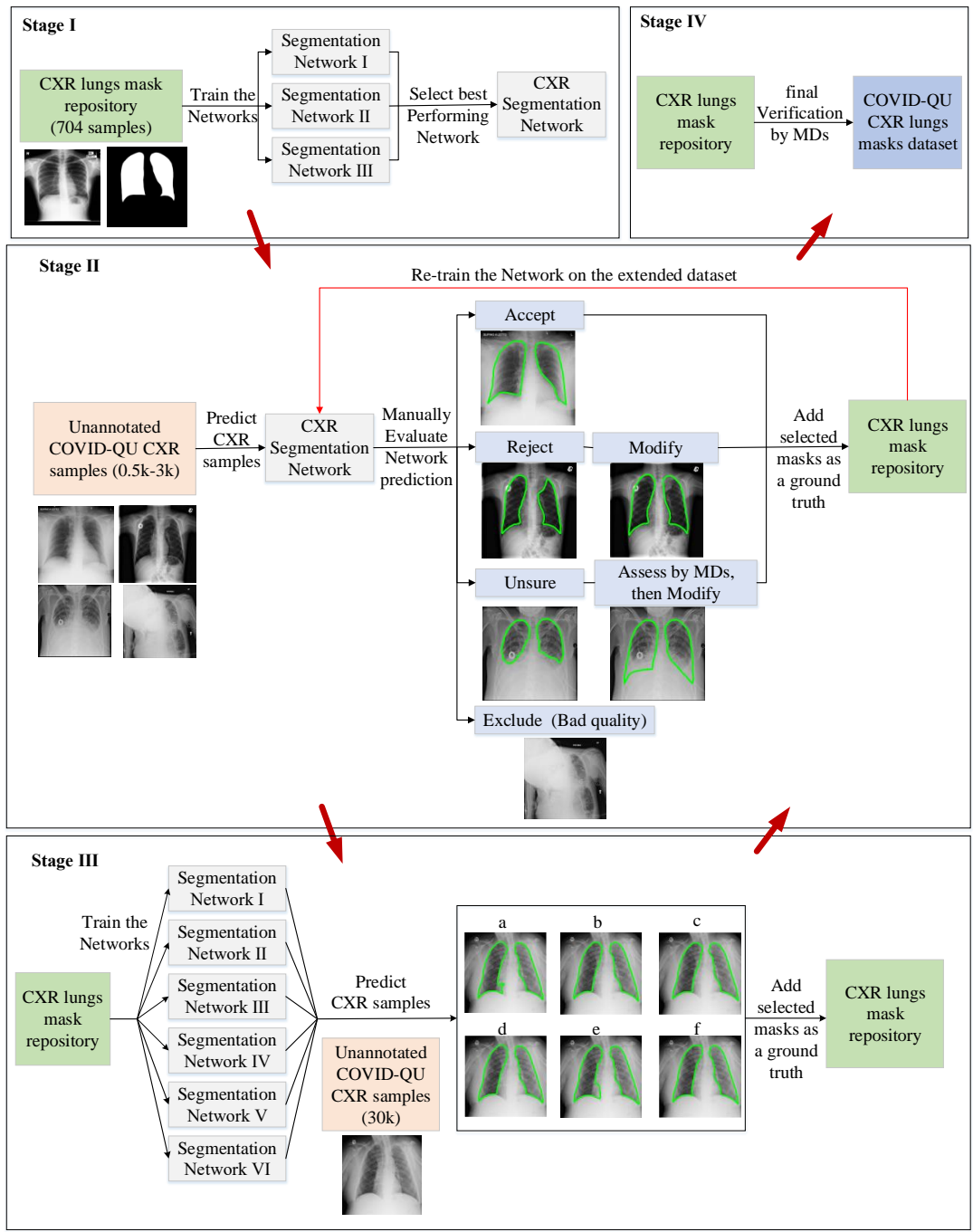

Figure 2. Collaborative human-machine approach to create ground-truth lung segmentation masks for COVID-QU CXR dataset. **Stage I**: Three segmentation networks are trained on a repository of 704 CXR lung segmentation masks, and the best network in terms of DSC is selected for the subsequent stages. **Stage II**: An iterative training is utilized to create lung masks for a subset of 3,000 CXR samples from COVID-QU dataset. Firstly, A subset of 500 samples is inferred by CXR segmentation model and the outputs are evaluated manually as accept, reject, modify, or exclude. Next, the modified masks are added to the lung repository and the network is re-trained on the extended dataset. These steps are repeated until generating ground-truth masks for the 3,000 CXR samples is completed. **Stage III**: six deep segmentation networks are trained using the 3,000 ground-truth masks generated in the previous stage. The trained networks are used to predict segmentation masks for the rest of the COVID-QU dataset (30,900 images). **Stage IV**: a final verification is performed by MDs on randomly selected 6,788 CXR samples (20% of the full dataset) that well presents the diversity of the COVID-QU dataset.

# 3 Methods

In this section, we describe the proposed unified approach for lung segmentation and COVID-19 localization with infection quantification from the CXR images. As the schematic representation of the pipeline of the proposed COVID-19 recognition system is shown in Figure 3, a binary lung mask is first generated from the input CXR image using the 1$^{st}$ encoder-decoder (E-D) CNN. In parallel, the input CXR is fed to the 2$^{nd}$ E-D CNN to generate COVID-19 infection masks. Then, the generated lung and infection masks are superimposed with the CXR image to localize and quantify COVID-19 infected lung regions. Finally, the generated infection mask is used to detect COVID-19 positive cases from COVID-19 negative cases. We will detail each process next.

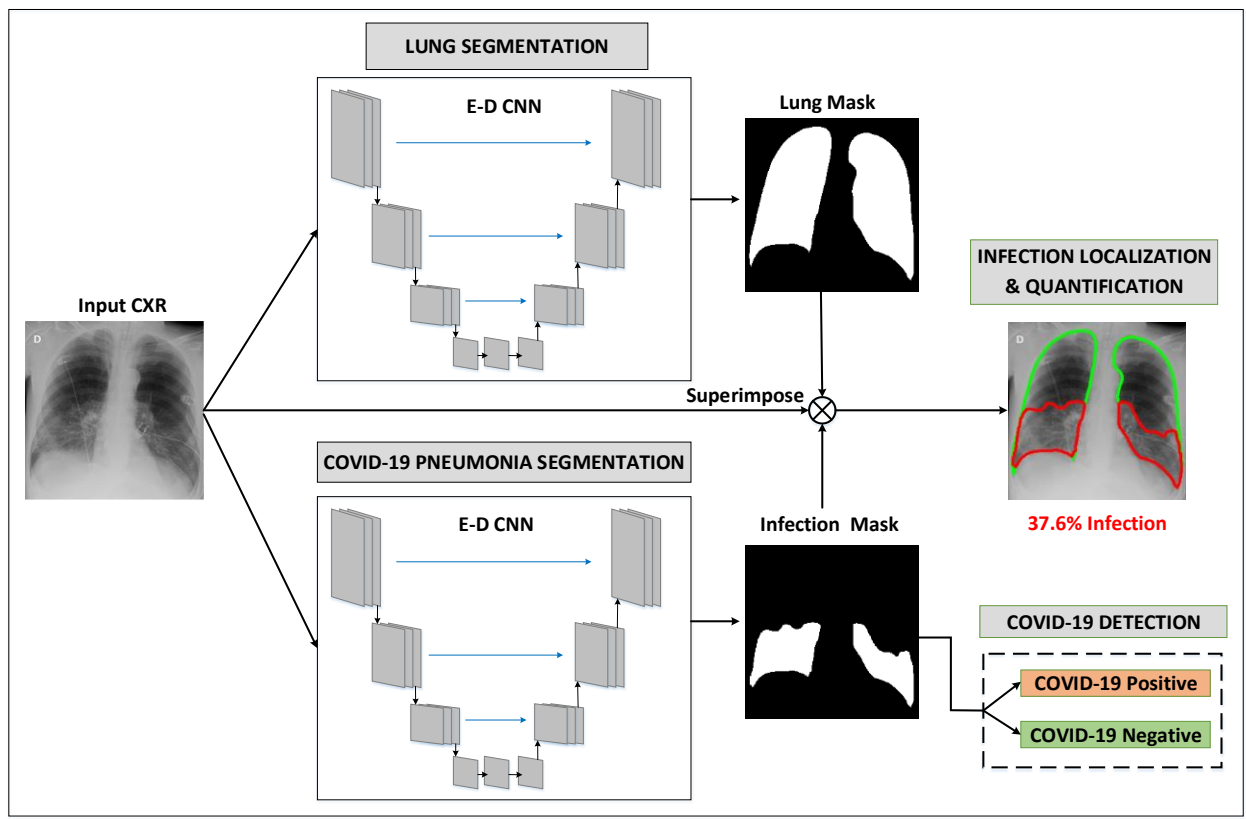

Figure 3. Schematic representation of the pipeline of the proposed system. The input CXR image is fed to two ED-CNNs in parallel, to generate two binary masks: lung, and COVID-19 infection masks. Next, the generated masks are superimposed with the CXR image to localize and quantify COVID-19 infected lung regions. Finally, the generated infection mask is used to detect COVID-19 positive cases from COVID-19 negative cases.

## 3.1  Network models for lung and COVID-19 infection segmentation

Lung and COVID-19 infection segmentation were performed on CXR images using three state-of-the-art deep E-D CNNs: U-Net [50], U-Net++ [51], and FPN [52], with different backbone (encoder) models using the variants of ResNet [53], DenseNet [54], and InceptionV4 [55] networks. Five variants of the backbone models were considered starting from shallow to deep structures: ResNet18, ResNet50, DenseNet121, DenseNet161, and InceptionV4.

The deployed encoder-decoder blocks provide a firm segmentation model that captures the context in the contracting path and empowers precise localization by the expanding path. The U-Net architecture has a classical decoder part that is symmetric to the encoder part, where max-pooling operations are replaced with up-sampling operations. Besides, high-resolution features from the encoder path are merged with the up-sampled output from the corresponding decoder path through skip connection. On the other hand, the U-Net++ is a recent implementation that has further developed the decoder block. The encoder and decoder blocks are connected through a series of nested dense convolutional blocks. This ensures a firm bridge between the encoder and decoder parts of the network, where information can be transferred to the final layers more intensively compared to the conventional U-Net. Both U-Net and U-Net++ architectures utilize 1×1 convolution to map the output from the last decoding block to two-channel feature maps, where a pixel-wise SoftMax activation function is applied to map each pixel into a binary class of background or lung for Lung segmentation task, and background or lesion for infection segmentation task. In contrast, the FPN employs the encoder-decoder as a pyramidal hierarchy by generating prediction masks at each spatial level of the decoder path. All predicted feature maps are upsampled to the same size, concatenated, convolved with a 3×3 convolutional filter, and then SoftMax activation is applied to generate the final prediction mask.

To ensure efficient training and faster convergence, transfer learning was utilized on the encoder side of the segmentation networks by initializing the convolutional layers with ImageNet weights.

### 3.1.1 Segmentation loss function

The cross-entropy (CE) loss is used as the cost function for the segmentation networks:

$$CE = -\frac{1}{K}\sum_k \sum_c y_k \log(p(x_k)) \tag{1}$$

where $x_k$ denotes the $k^{th}$ pixel in the predicted segmentation mask, $p(x_k)$ denotes its SoftMax probability, $y_k$ is a binary random variable getting 1 if $y_k = c$, otherwise 0, and $c$ denotes the class category, i.e., $c \in \{background, lung\}$ for the lung segmentation task, and $c \in \{background, lesion\}$ for the infection segmentation.

### 3.2 Post-processing

The predicted segmentation masks, $\hat{Y}$, by the segmentation models are defined as $\hat{Y}_{h,w} \in [0,1]$, where $h$ and $w$ represent the size of the image. In the post-processing step, binary segmentation masks are first generated by thresholding with a fixed value of 0.5. The predicted pixels are classified as lung if $\hat{y} > 0.5$ for the lung segmentation task, while classified as COVID-19 infection if $\hat{y} > 0.5$ for the infection segmentation task. The binary lung masks are further processed by hole filling and removal of small regions, <5% of the total positive predicted pixels. As a result, we increase the true-positives while minimizing the false-positives, non-lung regions that are falsely predicted as a lung. In contrast, infection masks are and operated with post-processed lung masks to ensure that the infection region falls within the lung area and remove the false positives outside the lung region.

*3.3 COVID-19 detection and quantification*

The detection of COVID-19 is performed based on the prediction maps generated by the infection segmentation network. Accordingly, a CXR image is classified as COVID-19 positive if at least one pixel of lung areas is predicted as COVID-19 infection, i.e., $p(x\_k) > 0.5$. Otherwise, the image is considered as COVID-19 negative, healthy people or patients with Non-COVID pneumonia. Furthermore, COVID-19 infection is quantified by computing the overall percentage of infected lungs. Equivalently, the sum of predicted infection pixels over the sum of predicted lung pixels. In addition, the infection percentage of each lung is computed, enabling doctors to assess the progression of COVID-19 for each lung individually.

*3.4 Experimental Setup*

The lung segmentation task was conducted over the benchmark COVID-QU dataset. In contrast, the infection segmentation and COVID-19 detection tasks were conducted over a subset of 2,913 CXR samples from the COVID-QU dataset with corresponding infection masks from the QaTa-Cov19 dataset [38], which was a sub-set of the COVID-QU dataset. The CXR images are resized to have a fixed dimension of 256×256 pixels to be used as the input for deep networks. All tasks were done with a 20% test set, an 80% train set, and five-fold cross-validation, where 20% of training data was used as a validation set to avoid overfitting. Table 1 summarizes the number of images per class used for training, validation, and testing.

Table 1. Number of mages per class and per fold used for lung segmentation, infection segmentation, and COVID-19 detection tasks

| Dataset Name | Task | Class | | # of Samples | Training Samples | Validation Samples | Test Samples |
|---|---|---|---|---|---|---|---|
| COVID-QU dataset | Lung Segmentation | COVID-19 | | 11,956 | 7,658 | 1,903 | 2,395 |
| | | Non-COVID | | 11,263 | 7,208 | 1,802 | 2,253 |
| | | Normal | | 10,701 | 6,849 | 1,712 | 2,140 |
| | | Total | | 33,920 | 21,715 | 5,417 | 6,788 |
| COVID-QU and QaTa-Cov19 [38] datasets | Infection Segmentation and COVID-19 Detection | COVID-19 positive | | 2,913 | 1,864 | 466 | 583 |
| | | COVID-19 negative | Non-COVID | 1,457 | 932 | 233 | 292 |
| | | | Normal | 1,456 | 932 | 233 | 291 |
| | | Total | | 5,826 | 3,728 | 932 | 1,166 |

Quantitative evaluations of the proposed approach are performed for the lung segmentation, infection segmentation, and COVID-19 detection tasks. The segmentation tasks are evaluated on the pixel-level, where the foreground (lung or infected region) was considered as the positive class and background as the negative class. For the COVID-19 detection task, the performance was computed per CXR sample, where X-rays with COVID-19 infection were considered as the positive class and X-rays of healthy people or patients with Non-COVID pneumonia were considered as the negative class.

The performance of deep CNNs was assessed using different evaluation metrics with 95% confidence intervals (CIs). Accordingly, CI for each evaluation metric was computed as follows:

$$r = z\sqrt{metric(1 - metric)/N} \qquad (2)$$

where, $N$ is the number of test samples, and $z$ is the level of significance that is 1.96 for 95% CI.

### 3.4.1 Segmentation Evaluation Metrics

The performance of the lung and lesion segmentation networks were evaluated using three evaluation metrics which are Accuracy, Intersection over Union (IoU), and Dice Similarity Coefficient (DSC):

$$Accuracy = \frac{TP + TN}{TP + TN + FP + FN} \qquad (3)$$

where *accuracy* is the ratio of the correctly classified pixels among the image pixels. *TP, TN, FP, FN* represent the true positive, true negative, false positive, and false negative, respectively.

$$Intersection\ over\ Union\ (IoU) = \frac{TP}{TP + FP + FN} \qquad (4)$$

$$Dice\ Similarity\ Coefficient\ (DSC) = \frac{2TP}{2TP + FP + FN} \tag{5}$$

where, both $IoU$ and $DSC$ are statistical measures of spatial overlap between the binary ground-truth segmentation mask and the predicted segmentation mask, while the main difference is that $DSC$ considers double weight for $TP$ pixels (true lung/lesion predictions) compared to $IoU$.

### 3.4.2 COVID-19 Detection Evaluation Metrics

The performance of the COVID-19 detection scheme was assessed using five evaluation metrics: Accuracy, Precision, Sensitivity, F1-score, and Specificity.

$$Precision = \frac{TP}{TP + FP} \tag{6}$$

where $precision$ is the rate of correctly classified positive class CXR samples among all the samples classified as positive samples.

$$Sensitivity = \frac{TP}{TP + FN} \tag{7}$$

where $sensitivity$ is the rate of correctly predicted positive samples in the positive class samples,

$$F1 = 2\frac{Precision \times Sensitivity}{Precision + Sensitivity} \tag{8}$$

where $F1$ is the harmonic average of precision and sensitivity.

$$Specificity = \frac{TN}{TN + FP} \tag{9}$$

where $specificity$ is the $sensitivity$ of the negative class samples.

PyTorch [56] library with Python 3.7 was used to train and evaluate the deep CNN networks, with an 8-GB NVIDIA GeForce GTX 1080 GPU card. Adam optimizer was used, with the initial learning rate, $\alpha = 10^{-4}$, momentum updates, $\beta_1 = 0.9$ and $\beta_2 = 0.999$, an adaptive learning rate which decreases the learning parameter by a factor of 5 if validation loss did not improve for 3 consecutive epochs, early stopping criterion of 8 epochs, where training stops if validation loss did not improve for 8 consecutive epochs, and mini-batch size of 4 images with 40 back propagation epochs.

## 4 Experimental Results

In this section, both numerical and qualitative results are reported with an extensive set of comparative evaluations for lung segmentation, infection segmentation, and COVID-19 detection tasks.

*4.1 Results - Lung Segmentation*

The performance of the lung segmentation models over the test (unseen) set is tabulated in Table 2. Each model was evaluated with five different encoder structures. For all models, it was observed that DenseNet encoders achieve the top-segmentation performances as they can share pieces of collective knowledge by densely connecting convolutional layers to their subsequent layers, therefore, preserving the information coming from the earlier layer through the output layer. The FPN model with DenseNet121 encoder holds the leading performance with 96.11% IoU, and 97.99% DSC.

Table 2. Performance metrics (%) for lung region and COVID-19 infected region segmentation computed over test (unseen) set with three network models and five encoder architectures.

| Task | Model | Encoder | Accuracy | IoU | DSC |
|---|---|---|---|---|---|
| Lung Segmentation | U-Net | ResNet18 | 99.07 ± 0.23 | 95.91 ± 0.47 | 97.88 ± 0.34 |
| | | ResNet50 | 99.08 ± 0.23 | 95.93 ± 0.47 | 97.89 ± 0.34 |
| | | **DenseNet121** | **99.1 ± 0.22** | **96.06 ± 0.46** | **97.96 ± 0.34** |
| | | DenseNet161 | 99.1 ± 0.22 | 96.02 ± 0.47 | 97.94 ± 0.34 |
| | | InceptionV4 | 99.07 ± 0.23 | 95.9 ± 0.47 | 97.88 ± 0.34 |
| | U-Net ++ | ResNet18 | 99.07 ± 0.23 | 95.9 ± 0.47 | 97.88 ± 0.34 |
| | | ResNet50 | 99.1 ± 0.22 | 96.04 ± 0.46 | 97.95 ± 0.34 |
| | | **DenseNet121** | **99.11 ± 0.22** | **96.1 ± 0.46** | **97.98 ± 0.33** |
| | | DenseNet161 | 99.09 ± 0.23 | 95.98 ± 0.47 | 97.92 ± 0.34 |
| | | InceptionV4 | 99.08 ± 0.23 | 95.96 ± 0.47 | 97.91 ± 0.34 |
| | FPN | ResNet18 | 99.06 ± 0.23 | 95.86 ± 0.47 | 97.86 ± 0.34 |
| | | ResNet50 | 99.07 ± 0.23 | 95.91 ± 0.47 | 97.88 ± 0.34 |
| | | **DenseNet121** | **99.12 ± 0.22** | **96.11 ± 0.46** | **97.99 ± 0.33** |
| | | DenseNet161 | 99.09 ± 0.23 | 96.01 ± 0.47 | 97.94 ± 0.34 |
| | | InceptionV4 | 99.07 ± 0.23 | 95.92 ± 0.47 | 97.89 ± 0.34 |
| Infection Segmentation | U-Net | ResNet18 | **98.02 ± 0.8** | **82.92 ± 2.16** | **88.1 ± 1.86** |
| | | ResNet50 | 97.84 ± 0.83 | 81.73 ± 2.22 | 87.02 ± 1.93 |
| | | DenseNet121 | 97.98 ± 0.81 | 82.53 ± 2.18 | 87.74 ± 1.88 |
| | | DenseNet161 | 97.86 ± 0.83 | 81.95 ± 2.21 | 87.19 ± 1.92 |
| | | InceptionV4 | 97.98 ± 0.81 | 82.03 ± 2.2 | 87.11 ± 1.92 |
| | U-Net ++ | ResNet18 | 97.9 ± 0.82 | 82.9 ± 2.16 | 88.06 ± 1.86 |
| | | ResNet50 | 97.93 ± 0.82 | 82.59 ± 2.18 | 87.78 ± 1.88 |
| | | **DenseNet121** | **97.97 ± 0.81** | **83.05 ± 2.15** | **88.21 ± 1.85** |
| | | DenseNet161 | 97.95 ± 0.81 | 81.55 ± 2.23 | 86.66 ± 1.95 |
| | | InceptionV4 | 97.9 ± 0.82 | 81.13 ± 2.25 | 86.22 ± 1.98 |
| | FPN | ResNet18 | 97.84 ± 0.83 | 81.9 ± 2.21 | 87.25 ± 1.91 |
| | | ResNet50 | 97.84 ± 0.83 | 80.83 ± 2.26 | 86.25 ± 1.98 |
| | | DenseNet121 | 97.99 ± 0.81 | 82.55 ± 2.18 | 87.71 ± 1.88 |
| | | DenseNet161 | 97.95 ± 0.81 | 81.89 ± 2.21 | 87.08 ± 1.93 |
| | | **InceptionV4** | **97.99 ± 0.81** | **83.08 ± 2.15** | **88.13 ± 1.86** |

The outputs of the top three networks compared with the ground-truth are shown in Figure 4. An interesting observation is that the three networks can reliably segment lung regions not only for COVID-19 cases, but for non-COVID-19 pneumonia as well with different severity levels: mild, moderate, or severe. This elegant performance is empowered by the large COVID-QU dataset (over 33k samples), which encapsulates CXR samples with different quality, resolution, and SNR levels from COVID-19, non-COVID-19, and normal classes. Therefore, the benchmark dataset can help researchers to overcome the challenges and limitations faced, mainly in the lung segmentation phase for COVID-19 or other lung pathology problems. As most of the previous approaches were trained over Montgomery [35] and Shenzhen [36] CXR lung mask datasets, which comprise medium and high-quality X-ray images from normal and TB classes, the previous segmentation approaches were falling in unseen scenarios, such as severe infection or low-quality images [33].

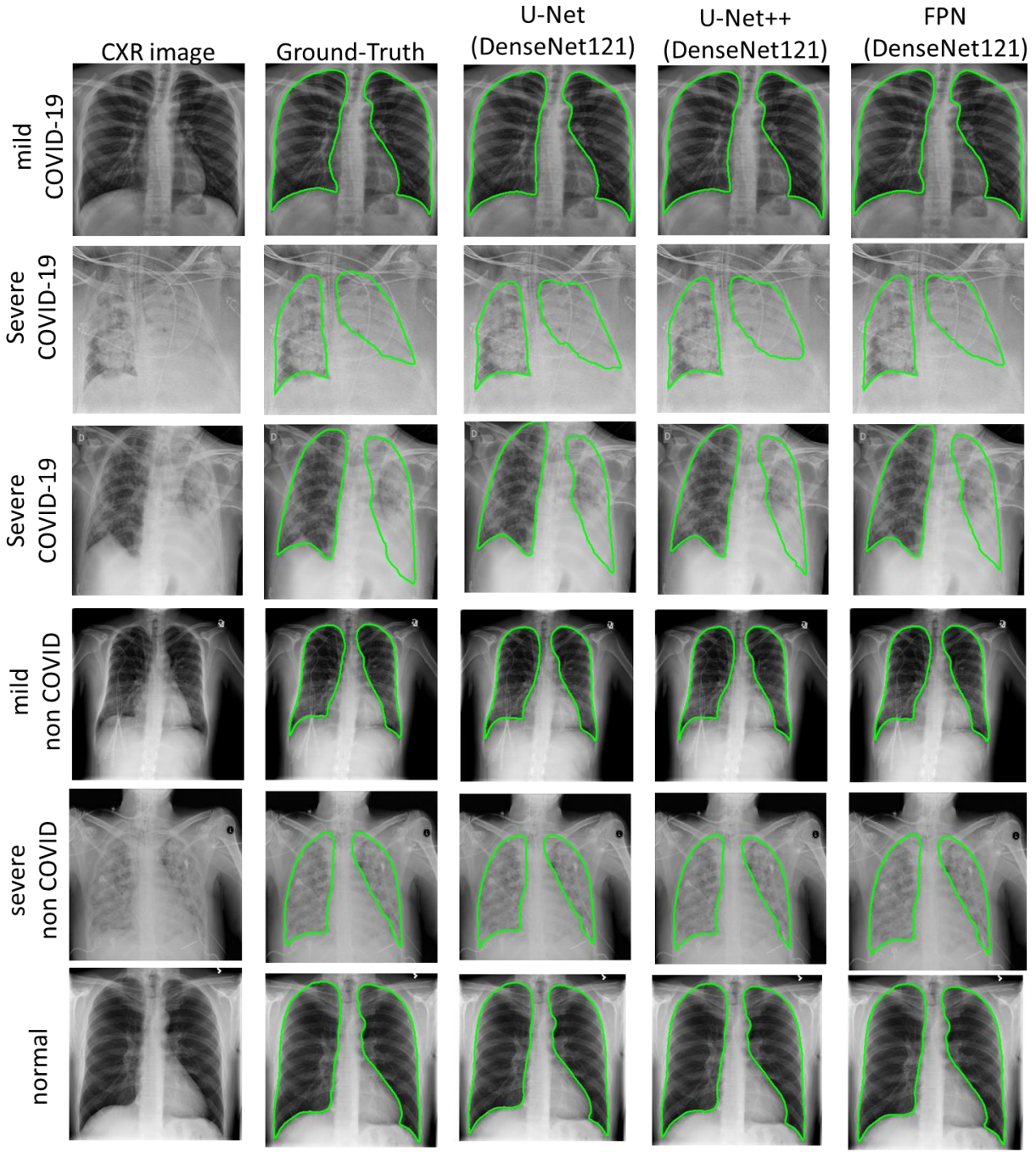

Figure 4. Qualitative evaluation of generated lung masks by top three networks. CXR image (1st column), ground truth (2nd column), and the lung masks of the top three networks (columns 3-5).

## 4.2 Results - Infection Segmentation

The infection segmentation model was first evaluated over two different configurations: cascaded and parallel segmentation. For the cascaded scheme, lung region was first segmented using the lung segmentation model; then the segmented CXR was fed to the infection segmentation model. While the plain

CXR was fed to both models independently for the parallel scheme. FPN model with DenseNet161 encoder was trained and evaluated on both schemes. The parallel scheme showed slightly better results with 87.08% DSC compared to 86.84% DSC for the cascaded scheme. Therefore, the parallel scheme was used as the main configuration for the remaining experiments. The performance of the infection segmentation models is presented in Table 2. U-Net++ model with DenseNet121 encoder showed the best performance with IoU and DSC values of 83.05% and 88.21%, respectively. Besides, the InceptionV4 encoder showed the highest performance among FPN models with 83.08% IoU and 88.13% DSC. In contrast, the shallowest encoder, ResNet18 presented the leading performance among U-Net models with IoU and DSC values of 82.92% and 88.1%, respectively.

Figure 5(a) shows the robustness of top-three networks to reliably segment COVID-19 infections of various shapes (small, medium, or large infection) with different severity levels (mild, moderate, severe, or critical infection). Figure 5(b) shows infection localization and severity grading of COVID-19 pneumonia for a 42-year female patient on the 1st day (admission to hospital), 2nd day, and 3rd day using the proposed COVID-19 recognition system.

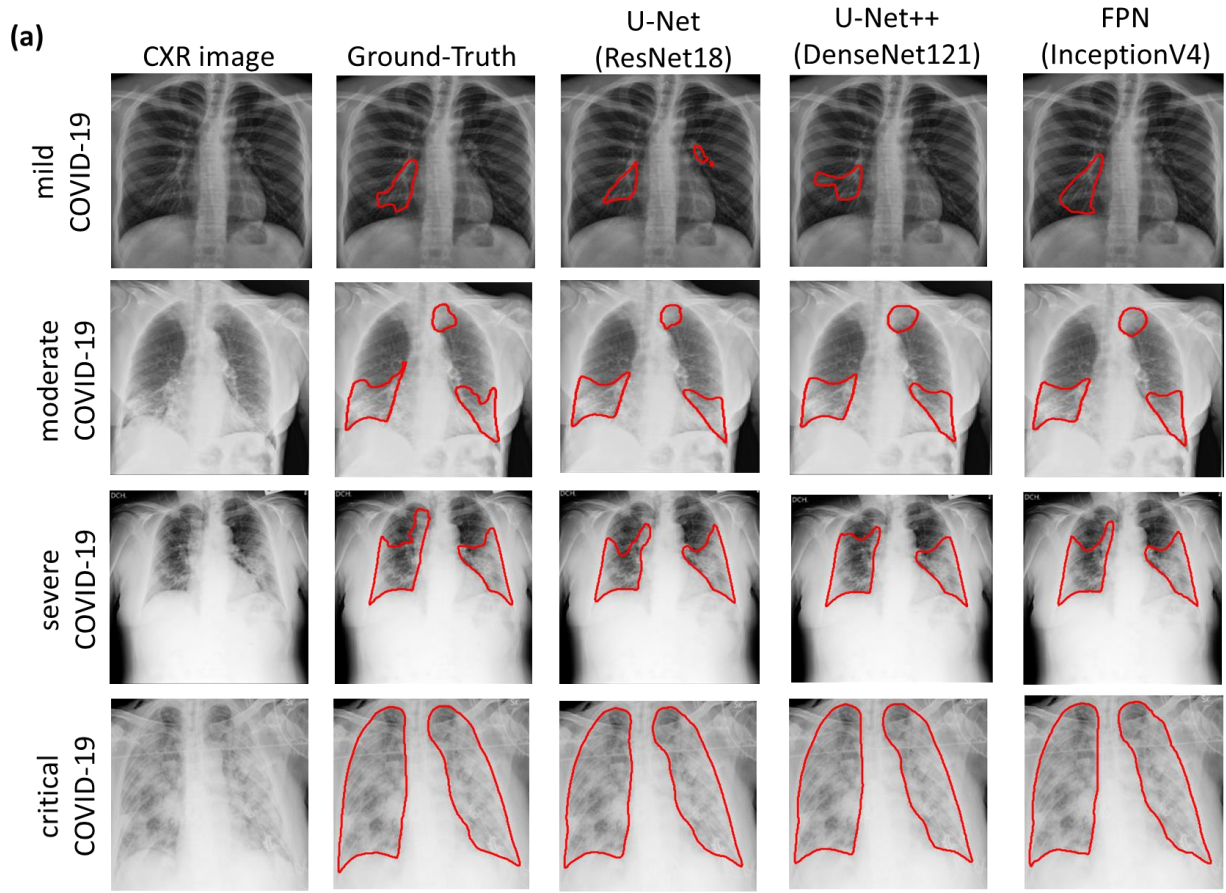

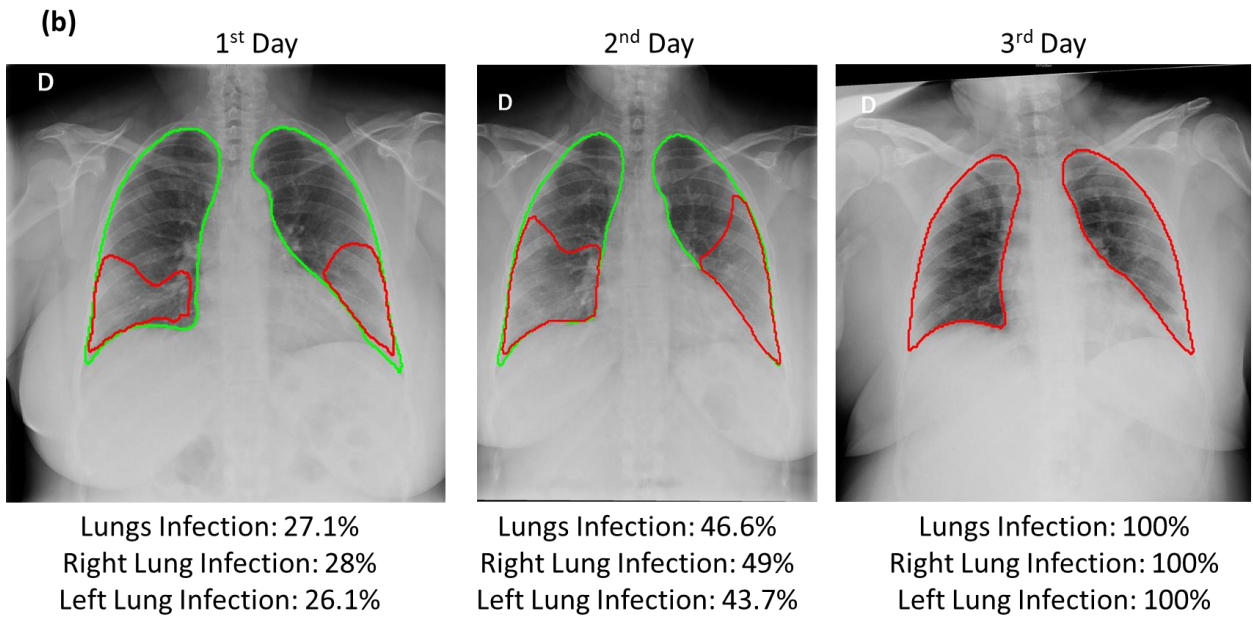

Figure 5. **(a)** Qualitative evaluation of generated infection masks by top three networks. CXR image (1st column), ground truth (2nd column), and the infection masks of the top three networks (columns 3-5). **(b)** Infection localization and severity grading of COVID-19 pneumonia for a 42-year female patient on the 1st, 2nd, and 3rd days using the proposed system.

*4.3   Results - COVID-19 Detection*

The performance of infection segmentation networks for COVID-19 detection from the CXR images is presented in Table 3. The sensitivity was considered as the primary metric for the detection task, as missing any COVID-19 positive case is critical. All the networks achieved high sensitivity values (>97%), where U-Net with DenseNet121 backbone and FPN with ResNet18 backbone achieved the best performance with a sensitivity of 99.66%. Similarly, all models showed high specificity values (>97%), where U-Net++ with ResNet18 backbone achieved the top performance with 100% specificity, indicating the absence of any false alarm.

Table 3. COVID-19 detection performance results (%) computed over test (unseen) set with three network models, and five encoder architectures.

| Model | Encoder | Accuracy | Precision | Sensitivity | F1-score | Specificity |
|---|---|---|---|---|---|---|
| U-Net | ResNet18 | 98.89 ± 0.6 | 99.14 ± 0.53 | 98.63 ± 0.67 | 98.88 ± 0.6 | 99.14 ± 0.53 |
|  | ResNet50 | 98.89 ± 0.6 | 98.47 ± 0.7 | 99.31 ± 0.48 | 98.89 ± 0.6 | 98.46 ± 0.71 |
|  | DenseNet121 | 98.8 ± 0.62 | 97.98 ± 0.81 | **99.66 ± 0.33** | 98.81 ± 0.62 | 97.94 ± 0.82 |
|  | DenseNet161 | 98.71 ± 0.65 | 97.97 ± 0.81 | 99.49 ± 0.41 | 98.72 ± 0.65 | 97.94 ± 0.82 |
|  | InceptionV4 | 98.03 ± 0.8 | 98.28 ± 0.75 | 97.77 ± 0.85 | 98.02 ± 0.8 | 98.28 ± 0.75 |
| U-Net ++ | ResNet18 | 99.23 ± 0.5 | 100 ± 0 | 98.46 ± 0.71 | 99.22 ± 0.5 | **100 ± 0** |
|  | ResNet50 | 99.14 ± 0.53 | 99.83 ± 0.24 | 98.46 ± 0.71 | 99.14 ± 0.53 | 99.83 ± 0.24 |
|  | DenseNet121 | 99.23 ± 0.5 | 99.14 ± 0.53 | 99.31 ± 0.48 | 99.22 ± 0.5 | 99.14 ± 0.53 |
|  | DenseNet161 | 98.2 ± 0.76 | 97.95 ± 0.81 | 98.46 ± 0.71 | 98.2 ± 0.76 | 97.94 ± 0.82 |
|  | InceptionV4 | 98.2 ± 0.76 | 98.45 ± 0.71 | 97.94 ± 0.82 | 98.19 ± 0.77 | 98.46 ± 0.71 |
| FPN | ResNet18 | 98.54 ± 0.69 | 97.48 ± 0.9 | **99.66 ± 0.33** | 98.56 ± 0.68 | 97.43 ± 0.91 |
|  | ResNet50 | 98.46 ± 0.71 | 98.46 ± 0.71 | 98.46 ± 0.71 | 98.46 ± 0.71 | 98.46 ± 0.71 |
|  | DenseNet121 | 98.97 ± 0.58 | 99.65 ± 0.34 | 98.28 ± 0.75 | 98.96 ± 0.58 | 99.66 ± 0.33 |
|  | DenseNet161 | 98.11 ± 0.78 | 97.3 ± 0.93 | 98.97 ± 0.58 | 98.13 ± 0.78 | 97.26 ± 0.94 |
|  | InceptionV4 | 99.23 ± 0.5 | 99.31 ± 0.48 | 99.14 ± 0.53 | 99.22 ± 0.5 | 99.31 ± 0.48 |

*4.4 Computational Complexity Analysis*

Table 4 compares the segmentation models in terms of inference time and the number of trainable parameters. The results present the running time per CXR sample. It can be seen that FPN and U-Net models are faster than U-Net ++ models, due to their shallow and close structures. FPN with ResNet18 encoder is the fastest network taking up to 5.74 ms per image. In contrast, the U-Net++ model is the slowest with the largest number of trainable parameters. The most computationally demanding model is UNet++ with InceptionV4 encoder with 59.35M trainable parameters. However, UNet++ with DenseNet161 encoder is the slowest, with an inference time of 48.62 ms, as it is the deepest model with 161 layers. Moreover, for systems with limited computational capabilities, where both lung and infection segmentation cannot be used in parallel, the two models can be used consecutively. This will double (×2) the inference time,

<100ms. However, we can still say that the full system can be used for real-time clinical applications as the overall inference time is still less than 100 ms. Therefore, multiple images can be processed within a second to take advantage of this state-of-the-art performance.

Table 4. The number of trainable parameters of the models with their inference time (ms) per CXR sample.

| Model | Encoder | Trainable parameters | Inference Time (ms) |
|---|---|---|---|
| U-Net | ResNet18 | 14.32M | 5.78 |
| | ResNet50 | 32.5M | 10.44 |
| | DenseNet121 | 13.60M | 22.86 |
| | DenseNet161 | 38.73M | 29.74 |
| | InceptionV4 | 48.79M | 26.53 |
| U-Net ++ | ResNet18 | 15.96M | 8.30 |
| | ResNet50 | 48.97M | 19.90 |
| | DenseNet121 | 30.06M | 25.13 |
| | DenseNet161 | 79.04M | 48.62 |
| | InceptionV4 | 59.35M | 32.53 |
| FPN | ResNet18 | 13.04M | 5.74 |
| | ResNet50 | 26.11M | 10.34 |
| | DenseNet121 | 9.29M | 22.68 |
| | DenseNet161 | 29.49M | 29.62 |
| | InceptionV4 | 43.57M | 26.08 |

## 5 Conclusion

Early identification and isolation of highly infectious COVID-19 cases play a vital role in preventing the spread of the virus. X-ray imaging is a low-cost, easily accessible, and fast method that can be an excellent alternative for conventional diagnostic methods such as RT-PCR and CT scans. Therefore, numerous studies proposed AI-based solutions for automatic and real-time detection of COVID-19. In general, these methods showed outstanding performance for early detection and diagnosis. However, they have used limited CXR repositories for evaluation with a small number, a few hundreds, of COVID-19 samples. Thus, the generalization of the achieved results on large cohort dataset is not guaranteed. In addition, they showed limited performance in infection localization and severity grading of COVID-19 pneumonia. In this study, we propose a robust system to segment the lung, detect, localize, and quantify COVID-19 infections from the CXR images. To accomplish this, we compiled the largest CXR dataset, COVID-QU, which consists of 11,956 COVID-19, 11,263 Non-COVID pneumonia, and 10,701 Normal images. Moreover, we constructed ground-truth lung segmentation masks for the benchmark dataset using an elegant collaborative human-machine approach, which can save valuable human labor time and minimize subjectivity in the annotation process. The publicly shared dataset will help researchers to investigate deep CNN on a comparatively larger dataset, which can provide more reliable solutions for COVID-19 and other lung pathology problems.

Extensive experiments on COVID-QU showed superior lung segmentation performance with 96.11% IoU and 97.99% DSC. Moreover, the proposed system proved reliable in localizing COVID-19 infection of various severity, achieving IoU and DSC values of 83.05% and 88.21%, respectively. Furthermore, an unprecedented COVID-19 detection performance was achieved with sensitivity and specificity values >99%. To the best of our knowledge, this is the first study that utilizes both lung and infection segmentation to detect, localize and quantify COVID-19 infection from X-ray images. Therefore, it can assist the medical doctors to better diagnose the severity of COVID-19 pneumonia and follow up the progression of the disease easily.

In the future, we plan to explore robust quantization and model compression techniques to further reduce the model complexity and accelerate the inference process, using the new generation of heterogeneous network models such as Self-Organized Operational Neural Networks [57, 58].

## Author contributions

Experiments were designed by AMT, MEHC and SK. Experiments were performed by AMT, AK, TR, YQ and UK. Data were compiled and created by AMT, AK, TR, YQ, UK, NI, KH, TH, SM and ME. Results were analysed by AMT, MEHC, SK, MSR, SAM, KH, and TH. The project is supervised by MEHC and SK. All the authors were involved in the interpretation of data and paper writing and revision of the article.

## Funding

Qatar University COVID19 Emergency Response Grant (QUERG-CENG-2020-1) provided the support for the work and the claims made herein are solely the responsibility of the authors.

## Competing Interest

The authors report no declarations of interest.